# Predicting User Performance and Bitcoin Price Using Block Chain Transaction Network

Afshin Babveyh
Stanford University
afshinb@stanford.edu

Sadegh Ebrahimi
Stanford University
sadegh@stanford.edu

**Introduction**

Bitcoin is a form of crypto currency introduced by Satoshi Nakamoto in 2009. Even though it only received interest from the software community at the beginning, today Bitcoin has a market cap of 5 billion dollars. Bitcoin offers a certain layer of anonymity to its users. Everyone can participate in the Bitcoin network without any registration process. Bitcoin transactions use a pair of elliptic curve cryptographic keys to redeem and spend the money. The money is held in the public key, which is also called address. Each public key has a corresponding secret key, known only to the owner of the account. The user can redeem the money in an address using his/her secret key [1]. The best practice is to use new keys for every new transaction to retain the user anonymity. If an address is used more than once, it would reveal some information about the users involved in the transaction. It could also reveal information about other users who were not part of this transaction, but simply had a transaction with one of the users involved in it.

The most popular type of transaction in Bitcoin is pay to public hash. Each transaction has an input field and an output field. The input field specifies the addresses, which contain the funds. The output address specifies the destination address. The funds specified in the input are equal to the funds in the output. For example if Alice owns two addresses A1, A2 each holding 5 Bitcoin, and wants to deposit 7 Bitcoin to Bob with address B1, the transaction will have A1 and A2 as input. It will have B1 with 7 Bitcoins, and A3 with 3 Bitcoins as the output. Alice controls A3, which is also called change address. There are several heuristic methods to cluster addresses into users. As an example, one could argue that all the addresses in the input field belong to the same entity which is used in creating the dataset on the course website. It is easy to see that if an address is used more than once, it would reveal information about anyone that ever had a transaction with it. The best practice in Bitcoin is to use a new address for every new transaction [2].

This work is organized as follows. In the first section we review the prior work and we have obtained our data. Next, we will look at address reuse in the Bitcoin network. We show that a great portion of users reuse their addresses which could enable us to cluster the addresses and attribute them to single users. Next, we will categorize the nodes based on their role in the network as a customer or seller. Finally, we do a study of nodes and network performance.

**Prior Work**

[3] discusses the problem of anonymity in the Bitcoin network. This work links public keys to real people by using the data available in online forums. They have demonstrated that

using the page rank algorithm would reveal nodes with many transactions such as Silk Road address.

[4] discusses methods to find dominant patterns in a graph. Their algorithm scans the graph for all n-node possible sub-graphs. The occurrence of a sub-graph is compared with its occurrence in a random graph. Network motifs are the patterns whose occurrence is very larger as compared to a random graph. We use the idea in the paper to assign a score and find a threshold for the Bitcoin graph by comparing it to a random graph. While it would be interesting to investigate the motifs in the graph, the enormous size of graph makes it impossible to run any algorithm worse than O(n).

**Data Collection**

We used two sources of data. The first one is the data publicly available at [5] as suggested by the course website. This dataset has already clustered the public key addresses into user nodes, which makes the analysis of the network easier. The problem with this dataset is that it dates back to April 2013. Our second source of data was the actual Bitcoin block data available on the its peer to peer network. Near 30GBs of the block chain data were downloaded (up to December 2014). We then used our parsing script [6] to extract the information from the binary block chain files. We have used the parser to extract the public keys (addresses) used in the transactions for investigating address reuse. For the rest of this work we have used the file available on the course website which clusters several addresses to a single user. The heuristic used to cluster the addresses assumes all the addresses in the input of a transaction belong to the same user entity. Although, this heuristic is not completely valid as it neglects cases such as coin mixing it provides a strong framework for the analysis of the network.

**Address Reuse**

Bitcoin network supports different kinds of transactions. The most popular scheme is pay to public key hash, where the destination address is hash of the destination's public key. The destination address can claim the funds using the secret key. Creating a pair of public and secret keys is a fast process and is a matter of generating random number. The best practice for anonymity in the Bitcoin network is to use a new key for every new transaction. However, there are some factors hindering this. For example, it would be very complicated for an ordinary user to frequently update an address he/she has listed on a website. Using an address (public key) to receive multiple payments could reveal some information about the customers. As a simple example, if we know that a given address belongs to merchant X, by looking at the Bitcoin ledger data, it is possible to tell who else has paid merchant X.
In this part we have investigated the practice used by Bitcoin users when it comes to using a new key for every transaction. To answer this question, we used 30GB of block chain data up to December 2014. We divided the transactions in 3 parts based on the transaction time. Each block contains 50 transaction files and around three hundred million transactions. For each time frame, the amount of address reuse in the network was calculated. Figure[1] shows a log-log plot of the distribution of address reuse in the network. As the figure shows, Bitcoin users tend to reuse their addresses. The probability of an address being used $r$ times is proportional to $p = r^{-2.5}$.

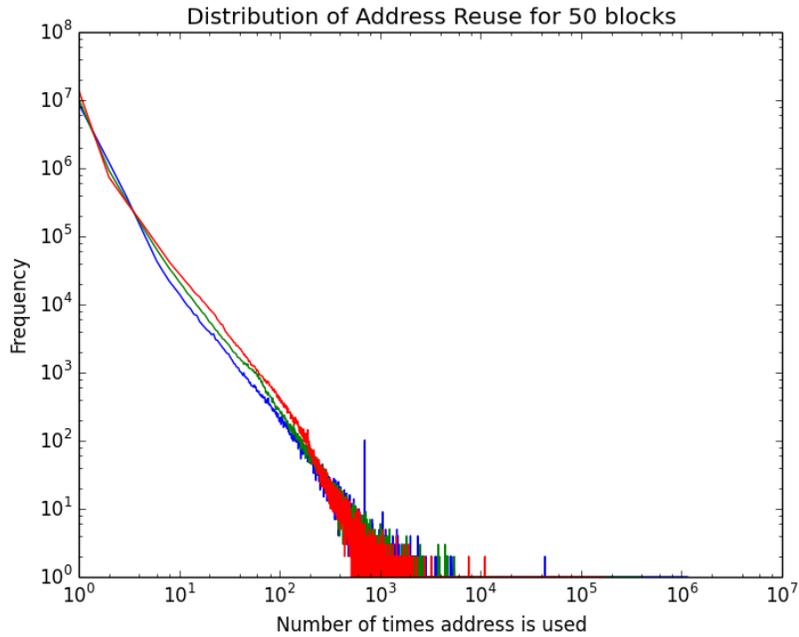

**Figure 1**

**Node Categorization**

One informative way to study dynamics of transactions in Bitcoin graph is to assign rolls to specific sub-populations of nodes and see how those sub-populations change size over time. We can do this sort of node categorization based on the transactional behaviors of the nodes such as the amount of money they spent or the amount of money they receive.
One simple categorization is to select nodes based on the difference between sum of their incoming transactions and sum of their outgoing transactions. If the total amount spent by a node is much more than the total amount received that means this node is producing Bitcoins (miner). On the other hand there are a group of nodes that receive much more than what they spend (collectors). These two groups of nodes are very important in shaping the dynamics of the whole graph because they are basically the source and destination of a majority of transactional flow in this network. So we are interested to see if there is any relation between the activity of these specific nodes and the rest of the network.
But in order to find these sub-populations we first must understand how much difference between incoming and outgoing money is significant enough to categorize the node as a miner or a collector. To find that we need to look into this difference in a random graph and see what is the expected difference if all the nodes were performing random transactions. To measure that we built a random Erdos-Renyi graph with the same number nodes (6M) and edges (16M) as the Bitcoin graph. For this purpose we used the random graph generator function to create a directed random Erdos-Renyi graph (GenRndGnm). Then we assigned a transaction value as an attribute to each edge. The transaction values were randomly sampled from Bitcoin transaction values in the original graph. So the transaction value distribution is expected to be the same in both original and random graph. Now that we have built this random transaction graph we can look at the distribution of different node properties (in-degree, in-degree value, out-degree, out-degree value, difference between incoming and outgoing money and etc.) and find thresholds to decide if a value is significantly big or not (given a specific p-value). For our work we set the p-value to 0.01 and any value in the top one percentile of the random graph is considered significantly big. This step provides us with a significance threshold for different metrics that some of them are mentioned in table 1. This means that for example if (outgoing value-incoming value) > 432.0260 for a node, the probability that this node is not a miner is less than 0.01. The advantage of setting a threshold this way is that our results are independent of the data and the distribution of the graph does not directly change the threshold.

**Table 1 Significance thresholds for node categorization (p<0.01)**

|         | Degree | Value    |
|---------|--------|----------|
| **In**      | 7      | 444.3681 |
| **Out**     | 7      | 445.1454 |
| **In - Out**| 5      | 431.5362 |
| **Out - In**| 5      | 432.0260 |

**Network evaluation**

Using these thresholds we can find the nodes that are miner or collector. Since some public keys might only be used for a short period of time, we considered a lifetime for each node that starts from the first transaction that each node has completed and ends on the last transaction. Therefore at any point in time, we have an active population of miner or collector nodes. Figure 2 shows the size of these populations over time and figure 3 is the difficulty of mining Bitcoins over the same period (downloaded from blockchain.info). Difficulty is a measure of how difficult it is to find a new block compared to the easiest it can ever be.

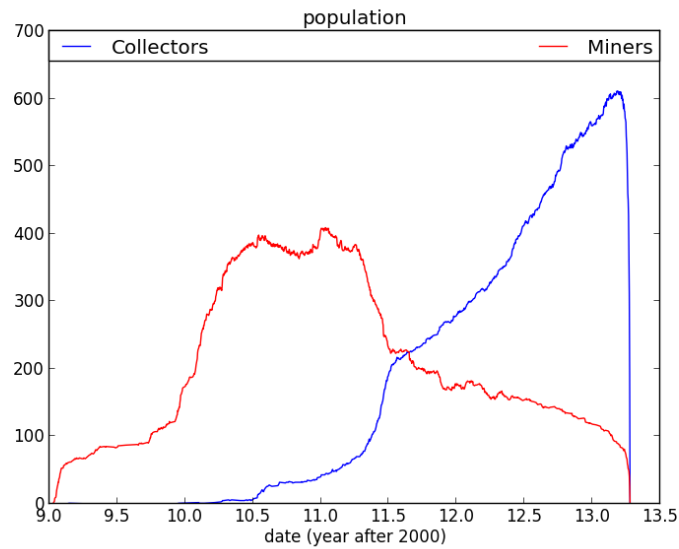

**Figure 2**

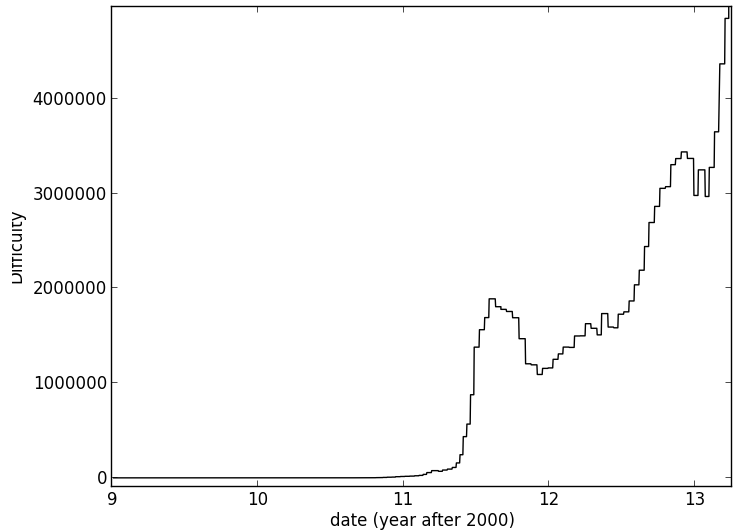

**Figure 3**

As we can clearly see in these two graphs the population of miner nodes has been increasing prior to 2011. But in the middle of 2011 there is a big increase in difficulty of mining Bitcoins. That is why in figure 2 we can see that population of miners start decreasing around that time. It is around this time that Bitcoin becomes valuable in the graph and we see a fast increase in population of collectors.

From figure 2 and 3 we can say that when difficulty of mining Bitcoin increased in 2011, many of the miners dropped out and more people started to collect Bitcoins and start investing in it. But how did this event affect the activity of other nodes in the graph?

In order to see this effect we studied the population of two bigger and more general categories of nodes: nodes that have significantly high ($p<0.01$) difference in their in-degree and out-degree assuming that someone who is a customer has most of its transactions as out going and the opposite for sellers. In figure 4 the population of active sellers and active customers has been shown. Although the population of both these groups in the network is more than 200,000 nodes, at any point in time only a few thousand of nodes are active.

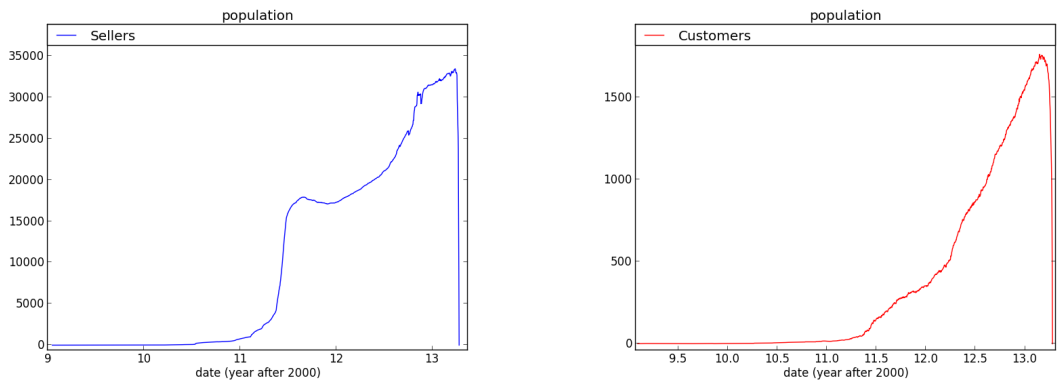

**Figure 4**

As we can see in figure 4, the rate of customer growth is increasing (the second derivative in time). So we could predict that in future probably the popularity of Bitcoin is increasing among normal population and customers.

We can then look at the correlation between ratio of customer population and seller population. This ratio is a good representative of supply and demand ratio. One then could look at Bitcoin as a valuable item that has a price and look at the correlation between supply and demand and price. Figure 5 shows this correlation between two values. Even though customer to seller ratio has a positive correlation with Bitcoin price (correlation coefficient=0.44) the relation is not linear. It seems as though you need to have enough increase in customer to seller ratio before it results in a stepwise jump in Bitcoin price. One could use such information to use in price prediction in Bitcoin market. Picture 6 shows how customer to seller ratio has changed during 2010 and 2013.

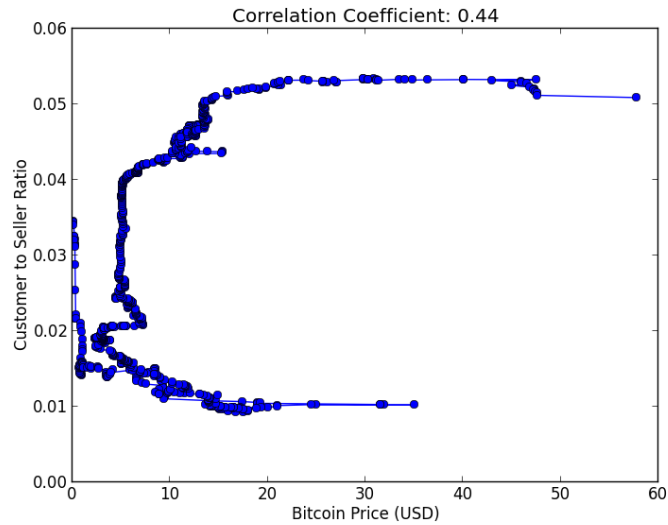

**Figure 5**

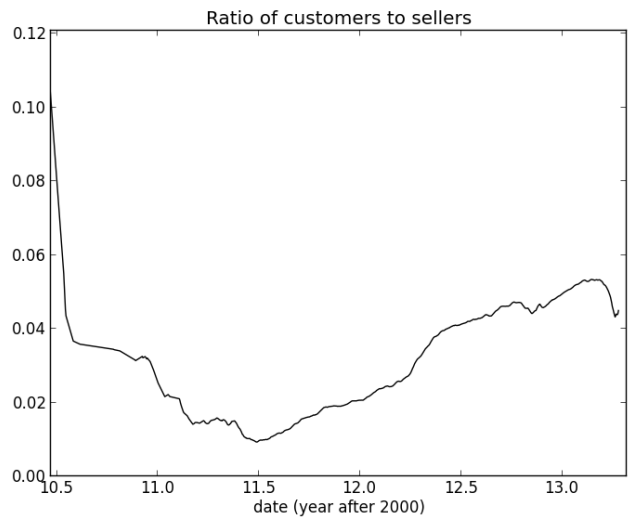

**Figure 6**

**Node Evaluation**

We used the list of sellers from previous section to investigate the correlation of centrality with money earned in the Bitcoin network. For this analysis we decided to exclude the customers as they are harder to track. This is because users are more likely to change their address after a purchase compared to when they receive money. A python script using snappy library was run to extract page rank, hubs and authorities for the graph. The resulting numbers were then used to investigate how these centrality measures are correlated with the money a user has earned. As seen in Figure 7, there is a correlation in the log-log space between the user's page rank and the amount of money he/she earns. The absolute amount of money is wide spread as seen in the graph. We speculate this is because different merchants offer different services, which could vary much in price.

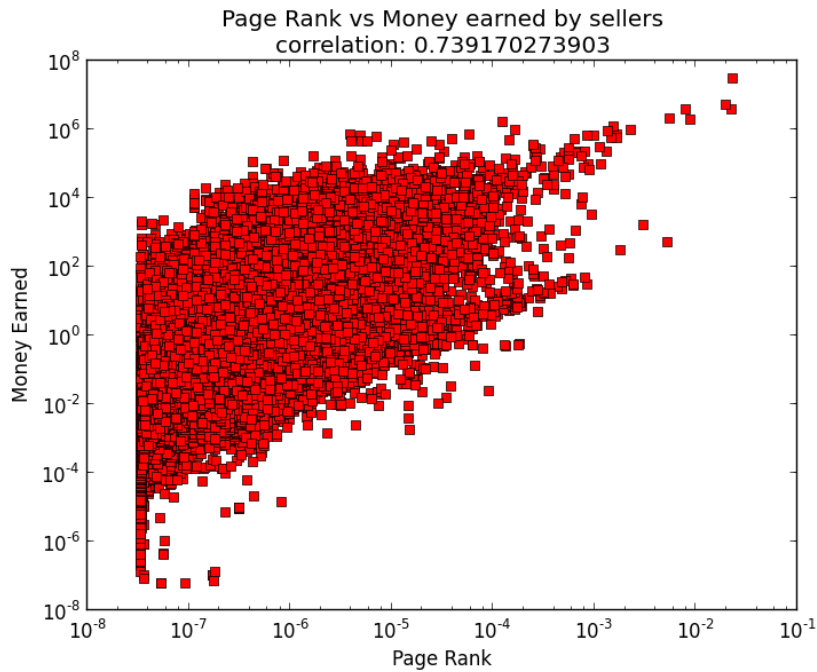

**Figure 7**

The correlation between the centrality figure and money earned by seller is outlined in Table 2. As the results in the table suggest, while there is a relevant correlation between a node's page rank and its success in the market, the hubs and authorities seem to be uncorrelated with the money earned by a seller.

Table 2

| Centrality Measure | Correlation with Money Earned |
|---|---|
| **Page Rank** | 0.74 |
| **Hubs** | 0.11 |
| **Authorities** | 0.98 |

The results indicate that successful sellers have a higher page rank.

**Conclusion**

This work addresses several questions about the Bitcoin network. We have shown that, the users in the Bitcoin network tend to reuse their addresses that would expose the users. We showed that the percentage of addresses reused *r* times in a transaction is given by $p = \alpha r^{-2.5}$. The users of the Bitcoin network are classified into sellers and customers based on a Z-score obtained from the study of a random graph. This kind of technique could be used to classify nodes and observe different group behaviors. For example in this work we showed that the ratio between customer population and seller population is very important factor and has a high correlation with bitcoin price. We have also shown that for the *sellers* the amount of money they earn is correlated with their page rank in the graph. For future work one could study other centrality measures to see how it is correlated with different properties of nodes. Also one could use these parameters as a feature vector to predict bitcoin value.